\documentclass{aastex63}
\shorttitle{Sample article}
\shortauthors{C.Xu et al.}

\begin{document}

\title{Refined Constraints on the Hubble Constant from Localized FRBs with Assessment of Systematic Effects}

\correspondingauthor{Yi Feng}
\email{yifeng@zhejianglab.org}

\author{Chenyuan Xu}
\affiliation{Research Center for Astronomical Computing, Zhejiang Laboratory, Hangzhou 311100, China}

\author{Yi Feng}
\affiliation{Research Center for Astronomical Computing, Zhejiang Laboratory, Hangzhou 311100, China}
\affiliation{Institute for Astronomy, School of Physics, Zhejiang University, Hangzhou 310027, China}

\author{Jiaying Xu}
\affiliation{Research Center for Astronomical Computing, Zhejiang Laboratory, Hangzhou 311100, China}





\begin{abstract}

The dispersion measure-redshift relation of fast radio bursts (FRBs) provides a valuable cosmological probe for constraining the Hubble constant, offering an independent measurement that could help resolve the ongoing Hubble tension. In this paper, we begin with a sample of 117 localized FRBs and use 95 of them to constrain $H_0$ to $71.28^{+1.90}_{-2.08}$ km s$^{-1}$ Mpc$^{-1}$ within the standard Lambda Cold Dark Matter ($\Lambda$CDM) model. The resulting statistical uncertainty is below 2.8\%, improving previous FRB-based measurements and highlighting the promise of larger future samples. Beyond statistical improvements, we note that different parameter choices have been adopted in previous studies and some results show discrepancies in $H_0$. To address this issue, we perform a systematic assessment of modeling uncertainties that can affect the inferred value of $H_0$, including Galactic electron density models, the contribution of the Galactic halo, outliers such as FRB~20190520B located in extreme environments, and the other parameter selections. We also discuss possible approaches to mitigate these sources of uncertainty, emphasizing both the challenges and prospects of using FRBs as reliable cosmological tools.

\end{abstract}



\section{Introduction} \label{sec_1}

As one of the most important parameters in the standard Lambda Cold Dark Matter ($\Lambda$CDM) model, the Hubble constant ($H_0$) is measured using the Cosmic Microwave Background (CMB) with the Planck satellite, yielding a value of $H_0 = 67.66 \pm 0.42$ km s$^{-1}$ Mpc$^{-1}$ \citep{refId0}. In contrast, measurements based on Cepheid-calibrated Type Ia Supernovae (SNIa), which serve as cosmic standard candles, give a value of $H_0 = 73.04 \pm 1.04$ km s$^{-1}$ Mpc$^{-1}$ \citep{Riess_2022}. The significant discrepancy between these two measurements has led to what is known as the Hubble tension. Fast radio bursts (FRBs), as a new cosmological probe, are believed to offer an independent means of measuring $H_0$ within the $\Lambda$CDM model, potentially helping to resolve this tension.

The study of FRBs, which are brief and intense pulses of radio emission originating from extragalactic sources and first reported in \citet{Lorimer_2007}, has emerged as a frontier in astrophysics and cosmology. FRBs are millisecond-duration, highly energetic radio transients of extragalactic origin, with frequencies typically spanning from 100 MHz to 8 GHz (\citealt{Pleunis_2021}, \citealt{Gajjar_2018}).
Despite over a decade of research since their first report, the physical origin of FRBs remains enigmatic.
The most plausible theoretical explanations \citep{2020Natur.587...45Z} suggest that repeating FRBs originate from magnetars, as demonstrated by the association of FRB~20200428A with the Galactic magnetar SGR~1935+2154 (\citealt{2020Natur.587...59B}, \citealt{Chime2020ABM}); in contrast, one-off FRBs are considered to result from compact object mergers.

Similarly to pulsars, FRB signals exhibit dispersion measures (DMs), characterized by frequency-dependent signal broadening. However, most FRBs arise from distant extragalactic sources, resulting in significantly larger dispersion measures than those of pulsars \citep{2022A&ARv..30....2P}. The dispersion measure of an FRB is generally given by the integral of electron number density along their propagation path, i.e., $\mathrm{DM}=\int_{0}^{d}n_e(l)dl$. In particular, a significant contribution to the FRB dispersion measure, denoted $\mathrm{DM_{IGM}}$, originates from the intergalactic medium. This property makes FRBs a promising cosmological probe for measuring the Hubble constant (\citealt{10.1093/mnras/stac2524}, \citealt{kalita2024fastradioburstsprobes}, \citealt{gao2024measuringhubbleconstantusing}), investigating dark energy (\citealt{Qiu_2022}, \citealt{2023SCPMA..6620412Z}), and locating the universe's missing baryons (\citealt{2020Natur.581..391M}, \citealt{connor2024gasrichcosmicweb}), among other applications. 

In \citet{2020Natur.581..391M}, a framework was established to constrain the uncertainty in $H_0$ by exploiting the correlation between redshift and dispersion measure (i.e., the $z$-DM relation or the Macquart relation), based on a sample of five FRBs with localized host galaxies. Subsequently, several studies have built upon this framework. For example, \citet{10.1093/mnras/stac2524} used 16 localized FRBs and 60 unlocalized FRBs to derive $H_0=73^{+12}_{-8}$ km s$^{-1}$ Mpc$^{-1}$, while \citet{Baptista_2024} employed 21 localized FRBs and 57 unlocalized FRBs, yielding $H_0=85.3^{+9.3}_{-8.1}$ km s$^{-1}$ Mpc$^{-1}$. \citet{kalita2024fastradioburstsprobes} utilized 64 localized FRBs and reported $H_0=83.53^{+2.32}_{-2.92}$ km s$^{-1}$ Mpc$^{-1}$, recently \citet{gao2024measuringhubbleconstantusing} and \citet{wang2025probingcosmology92localized} obtained $H_0=70.41^{+2.28}_{-2.34}$ km s$^{-1}$ Mpc$^{-1}$ and $H_0=69.04^{+2.30}_{-2.07}$ km s$^{-1}$ Mpc$^{-1}$ by 77 localized FRBs and 92 localized FRBs respectively.

As the sample size of FRB increases, the statistical uncertainty in $H_0$ has progressively decreased in recent studies. However, previous studies have used varying parameter settings, some yielding different results. To date, no studies have performed a comprehensive analysis of the sources of these differences, particularly the origins of systematic errors. 
Thus, we carry out the following investigations: (1) We start from a sample of 117 localized FRBs and select 95 to constrain $H_0$, achieving the smallest statistical uncertainty reported in relevant studies; (2) we comprehensively investigate systematic errors that can influence the result, thereby explaining the differences in $H_0$ reported in early studies.

\section{Methods}\label{sec_2}
Radio pulses experience dispersion as they travel through ionized media, leading to frequency-dependent time delays, with lower frequencies arriving later. The DM of an FRB can be measured using the time delay $\delta t$
between frequency $v_1$ and frequency $v_2$ (\citealt{Lorimer_2007}, \citealt{Deng_2014}, \citealt{2019A&ARv..27....4P}):
\begin{equation}
\delta t \simeq 4.15 \, \mathrm{s} \left[ \left( \frac{\nu_1}{\mathrm{GHz}} \right)^{-2} - \left( \frac{\nu_2}{\mathrm{GHz}} \right)^{-2} \right] \frac{\mathrm{DM}}{10^3 \, \mathrm{pc \, cm^{-3}}} .
\end{equation}

Physically, the DM of an FRB can be decomposed into four primary components \citep{2020Natur.581..391M}: 

\begin{equation}
\mathrm{DM}_{\mathrm{FRB}}(z) = \mathrm{DM}_{\mathrm{MW,ISM}} + \mathrm{DM}_{\mathrm{MW,halo}} + \mathrm{DM}_{\mathrm{IGM}}(z) + \frac{\mathrm{DM}_{\mathrm{host}}}{1+z} .
\label{equ_2}
\end{equation}

In Equation \ref{equ_2}, $\mathrm{DM_{MW,halo}}$ and $\mathrm{DM_{MW,ISM}}$ correspond to the contribution of the Galactic halo and the interstellar medium of the Milky Way, respectively. $\mathrm{DM_{host}}$ represents the local environment of the host galaxy and is scaled by $(1+z)^{-1}$. The intergalactic term, $\mathrm{DM}_{\mathrm{IGM}}(z)$, accounts for the dispersion accumulated along the line of sight through the intergalactic medium, and increases with redshift due to the longer path length.

Both $\mathrm{DM}_{\mathrm{IGM}}$ and $\mathrm{DM}_{\mathrm{host}}$ exhibit redshift dependence. For $\mathrm{DM}_{\mathrm{host}}$, the scaling factor $(1+z)^{-1}$ accounts for cosmological effects: it converts the host-frame dispersion into the observer frame by incorporating time dilation and redshifting of the radio pulse frequency \citep{2020Natur.581..391M}. For $\mathrm{DM}_{\mathrm{IGM}}(z)$, the mean value under the standard $\Lambda$CDM model has been derived by \citet{Deng_2014} and \citet{PhysRevD.89.107303}:

\begin{equation}
\langle \mathrm{DM_{IGM}} \rangle = \Omega_\mathrm{b} \frac{3 H_0 c}{8 \pi G m_p} \int_0^z \frac{(1 + z) f_{\mathrm{IGM}} \left[ \frac{3}{4} X_{e,\mathrm{H}}(z) + \frac{1}{8} X_{e,\mathrm{He}}(z) \right]}{\left[ \Omega_\mathrm{m} (1 + z)^3 + \Omega_\Lambda \right]^{1/2}} dz
 .
\label{equ_3}
\end{equation}

In Equation \ref{equ_3}, $\Omega_\mathrm{b}$, $\Omega_{\mathrm{m}}$ and $\Omega_\Lambda$ are the
densities of baryon, matter, and dark energy in units of $\rho_{c,0}=3H_0^2/8\pi G$, respectively. $f_{\mathrm{IGM}}$ is the baryon fraction that remains in the intergalactic medium from the Big Bang, 
a typical value adopted in previous studies is $\sim$0.844 (\citealt{Shull_2012}, \citealt{Zhang_2018}), with only a weak dependence on $z$. For the latest results, \citet{zhang2025probingcosmicbaryondistribution} provided an updated estimate of $f_{\mathrm{IGM}} = 0.865^{+0.101}_{-0.165}$ based on their recent simulations. On the observational side, \citet{connor2024gasrichcosmicweb} analyzed FRB samples and obtained a value of $f_{\mathrm{IGM}} = 0.93^{+0.04}_{-0.05}$, which is consistent with \citet{zhang2025probingcosmicbaryondistribution} within the quoted uncertainties. Note that $f_{\mathrm{IGM}}$ here refers to the total diffuse baryon fraction contributing to the observed DM, including both intergalactic medium and halo gas components.

\begin{figure*}
\gridline{\fig{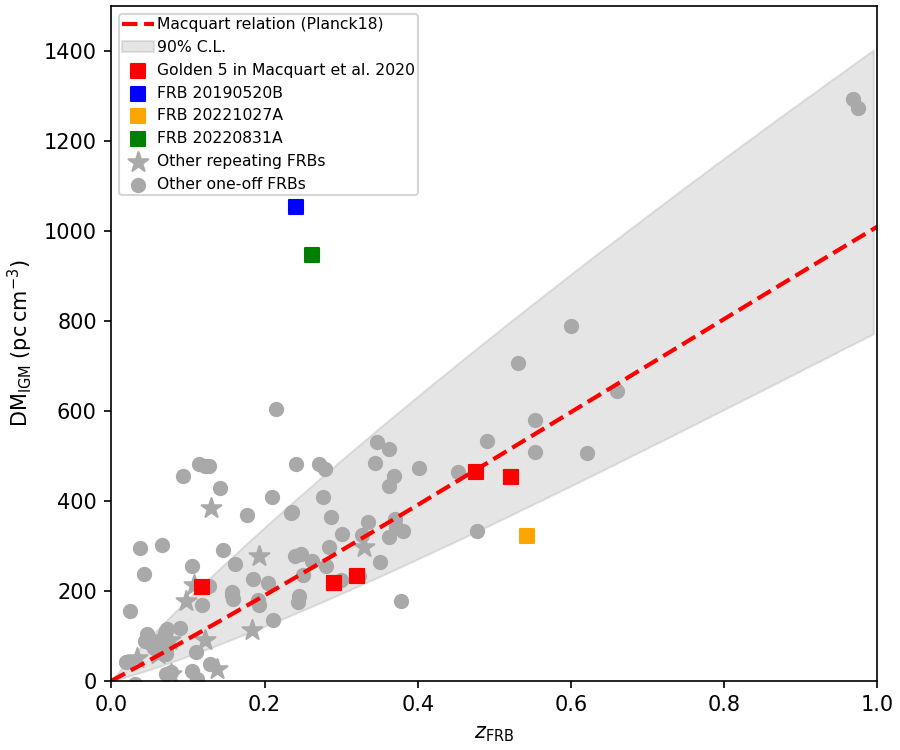}{0.48\textwidth}{(a)}
          \fig{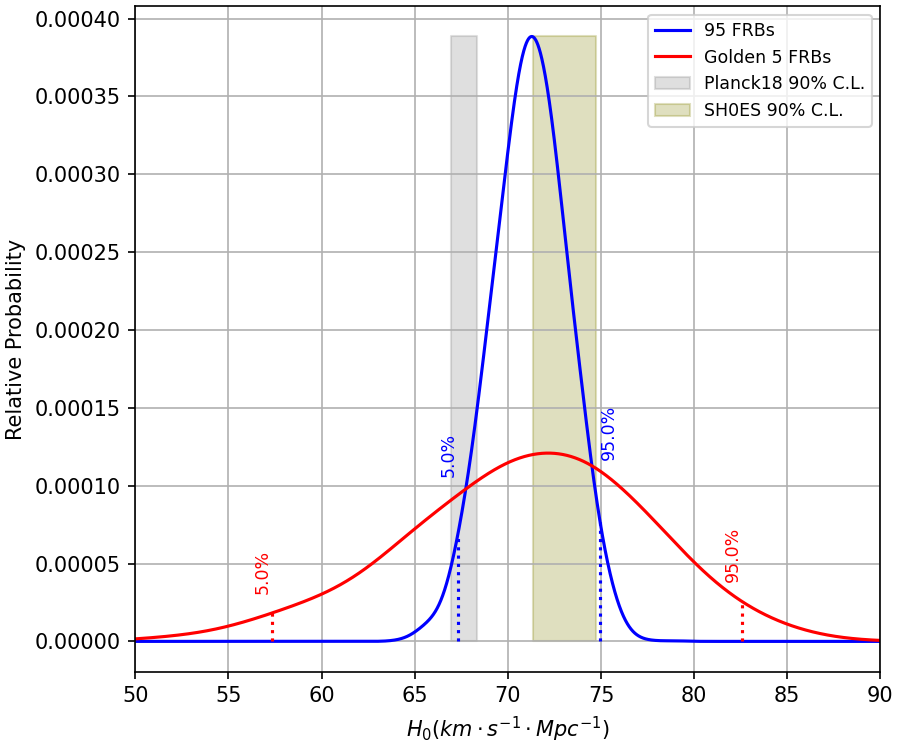}{0.48\textwidth}{(b)}
          }
\caption{(a) The Macquart relation of all FRBs used in this paper, $\mathrm{DM_{MW,ISM}}$ is calculated with NE2001, $\mathrm{DM_{MW,halo}}$ and $\mathrm{DM_{host}}$ are both assumed with 50 pc cm$^{-3}$, cosmology parameters are from Planck18 \citep{refId0} and $F$ is set to 0.32 by following \citet{2020Natur.581..391M}. (b) The $H_0$ posteriori distribution obtained by 95 FRBs in this paper using NE2001, yielding $H_0=71.28^{+1.90}_{-2.08}$ km s$^{-1}$ Mpc$^{-1}$. By comparison, the "Golden 5 FRBs" used in \citet{2020Natur.581..391M} yields $H_0=72.16^{+5.14}_{-7.58}$ km s$^{-1}$ Mpc$^{-1}$. The settings of MCMC are summarized in Table \ref{tab_1}.}
\label{fig_1}
\end{figure*}

In Equation \ref{equ_3}, $X_{e,\mathrm{H}}$ and $X_{e,\mathrm{He}}$ are hydrogen and helium ionization fractions, which can be considered as $X_{e,\mathrm{H}}\approx X_{e,\mathrm{He}}\approx1$ for $z<3$. In particular, a close approximation of the Macquart relation can be made as $\langle \mathrm{DM_{IGM}} \rangle \propto \Omega_\mathrm{b} H_0 f_{\mathrm{IGM}}z$ for $z<1$ (\citealt{2020Natur.581..391M}, \citealt{2022A&ARv..30....2P}), which is represented as a line in Figure \ref{fig_1}a. 

The deviation of $\mathrm{DM_{IGM}}$ from the mean value was originally modeled in \citet{McQuinn_2014}, then simplified in \citet{2019MNRAS.485..648P} and \citet{2020Natur.581..391M} with $\Delta\equiv \mathrm{DM_{IGM}}/\left \langle \mathrm{DM_{IGM}} \right \rangle $:

\begin{equation}
p_{\mathrm{DM_{IGM}}}(\Delta) = A \Delta^{-\beta} \exp \left[ - \frac{(\Delta^{-\alpha} - C_0)^2}{2 \alpha^2 \sigma_{\mathrm{IGM}}^2} \right], \quad \Delta > 0,
\label{equ_4}
\end{equation}
where $\alpha$ and $\beta$ are parameters that represent the slope of gas density profile of intervening halos, and determined as $\alpha\simeq 3, \beta\simeq 3$ in \citet{2020Natur.581..391M}. Parameter $A$ is for normalization, and $C_0$ is used to ensure the mean value is at $\langle \Delta \rangle=1$. $\sigma_{\mathrm{IGM}}$ is the fractional standard deviation of $\mathrm{DM_{IGM}}$, and can be approximated as $\sigma_{\mathrm{IGM}}=Fz^{-0.5}$ \citep{2020Natur.581..391M} for $z<1$. 
Note that earlier studies (\citealt{McQuinn_2014}, \citealt{2020Natur.581..391M}, \citealt{2022A&ARv..30....2P}) attribute the fluctuations in $\mathrm{DM}_{\mathrm{IGM}}$—characterized by $\sigma_{\mathrm{IGM}}$—to Poisson variations in the number of galaxy halos intersected along the line of sight, as well as to the compactness of the gas within these halos, assuming that the contributing electrons reside mainly in halos or their circumgalactic media.
In contrast, more recent works (\citealt{zhang2025probingcosmicbaryondistribution}, \citealt{Medlock_2025}) propose a different interpretation: the amplitude of $\sigma_{\mathrm{IGM}}$ is governed by baryonic feedback processes that expel gas from galaxy halos into the intergalactic medium, indicating that baryonic feedback may be stronger or more effective than previously assumed. While the physical interpretation of the fluctuation origin has shifted, the functional dependence remains consistent: a smaller value of $F$ corresponds to stronger baryonic feedback and a smaller $\sigma_{\mathrm{IGM}}$.

For the other contributions in Equation \ref{equ_2}:
\begin{itemize}
\item{$\mathrm{DM_{MW,ISM}}$: This term can be predicted by Galactic electron density model NE2001 \citep{2002astro.ph..7156C} or YMW16 \citep{Yao_2017}, both of which are built using galatic pulsars.}
\item{$\mathrm{DM_{MW,halo}}$: The exact value of $\mathrm{DM_{MW,halo}}$ is still uncertain. \citet{2019MNRAS.485..648P} estimated the range to be 50$\sim$80 pc cm$^{-3}$, and \citet{2020Natur.581..391M} chose a fix value of $\mathrm{DM_{MW,halo}}$=50 pc cm$^{-3}$. \citet{Yamasaki_2020} used X-ray data to predict $\mathrm{DM_{MW,halo}}$ with a mean value of 43 pc cm$^{-3}$, and their model YT2020 is adopted in recent FRB localization analysis \citep{amiri_2025}.}
\item{$\mathrm{DM_{host}}$: \citet{2020Natur.581..391M} employed a log-normal probability distribution to describe $\mathrm{DM_{host}}$, by which an asymmetric tail extending to extremely large values can be allowed:
\begin{equation}
p_{\mathrm{DM_{host}}}(\mathrm{DM_{host}}) = \frac{1}{\sqrt{2\pi \, \mathrm{DM_{host}} \, \sigma_{\mathrm{host}}}} \exp \left( - \frac{(\ln \mathrm{DM_{host}} - \mu_{\mathrm{host}})^2}{2 \sigma_{\mathrm{host}}^2} \right),
\label{equ_6}
\end{equation}
where the median is $e^{\mu_{\mathrm{host}}}$ and the variance is $e^{2\mu_{\mathrm{host}}} e^{{\sigma_{\mathrm{host}}}^2} \left( e^{{\sigma_{\mathrm{host}}}^2} - 1 \right)$. We note that the choice of Equation \ref{equ_6} has limited theoretical guarantees \citep{2020Natur.581..391M}, but the log-normal distribution fits well with the simulation results \citep{Zhang_2020}. 
}
\end{itemize}

With all the terms discussed above, the total likelihood $\mathcal{L}$ of a sample of $N$ FRBs can be built by multiplying the individual likelihoods for each FRB:

\begin{equation}
\mathcal{L} = \prod_{i=1}^{N_{\mathrm{FRB}}} \mathcal{L}_i,
\label{equ_7}
\end{equation}
\begin{equation}
\mathcal{L}_i=p_{\mathrm{IGM}}(\mathrm{DM_{i}-DM_{host,i}-DM_{MW,ISM,i}-DM_{MW,halo,i}} | \sigma_{\mathrm{IGM}}, H_0)\,p_{\mathrm{host}}(\mathrm{DM_{host,i}} | \mu_{host}, \sigma_{\mathrm{host}}).
\label{equ_8}
\end{equation}

Thus, the value of $H_0$ can be estimated using the Markov Chain Monte Carlo (MCMC). 

\section{Results}
With the likelihood built in Section \ref{sec_2}, we employ MCMC to estimate the posterior distribution of the Hubble constant $H_0$, treating it as the sole free parameter with a uniform prior $\mathcal{U}(50, 90)$. All other cosmological parameters, such as $\Omega_\mathrm{m}$ and $\Omega_\Lambda$, are fixed to the results of Planck18 \citep{refId0}. We use kernel density estimation (KDE) to smooth the posterior samples. The choice of a uniform prior $\mathcal{U}(50, 90)$ for $H_0$ is to avoid introducing any specific assumptions and to fully cover the ranges favored by both CMB-based and SH0ES local measurements. This treatment follows the approach originally adopted in \citet{2020Natur.581..391M}, which has also been followed by most subsequent studies using FRBs to constrain cosmological parameters (\citealt{10.1093/mnras/stac2524}, \citealt{Baptista_2024}, \citealt{kalita2024fastradioburstsprobes}, \citealt{wang2025probingcosmology92localized}).

Our results are shown in Figure \ref{fig_1}b and Table \ref{tab_1}, where we compare our results with those of the "Golden 5 FRBs" used in \citet{2020Natur.581..391M}, as well as several of the most recent studies. The detailed settings for building the likelihood are summarized in Table \ref{tab_1}, both for ours and the others. 

We explain our settings in more detail and will discuss these selections further in the following sections:
\begin{itemize}
\item $f_{\mathrm{IGM}}$: Earlier studies prefer the value of $\sim$0.844. More recent works, such as \citet{zhang2025probingcosmicbaryondistribution} and \citet{connor2024gasrichcosmicweb}, provide updated estimates based on simulations and observational data, respectively. Following the principle of relying on observational constraints whenever possible, we adopt the value $f_{\mathrm{IGM}} \simeq 0.93$ from \citet{connor2024gasrichcosmicweb}.
\item $\sigma_{\mathrm{IGM}}$: An approximation $\sigma_{\mathrm{IGM}}=Fz^{-0.5}$ was proposed in \citet{2020Natur.581..391M}, with a fiducial value of $F=0.32$ commonly adopted in subsequent studies (\citealt{James_2021}, \citealt{10.1093/mnras/stac2524}, \citealt{Fortunato_2023}, \citealt{Moroianu_2023}). \citet{Zhang_2021} fitted the IllustrisTNG simulation data and provided a table of $\sigma_{\mathrm{IGM}}$, $A$ and $C_0$, which are preferred by recent studies. However, we find that $C_0$ in \citet{Zhang_2021} does not hold $\langle \Delta \rangle=1$ in Equation \ref{equ_4}. Therefore, we follow the form $\sigma_{\mathrm{IGM}} = F z^{-0.5}$ and adopt the observationally constrained value $F = 10^{-0.75}$ from \citet{Baptista_2024}.
\item $\mu_{\mathrm{host}}$, $\sigma_{\mathrm{host}}$: Earlier studies fitted these two as free parameters, while recent studies tend to use fixed values to improve the constraining power on $H_0$. We adopt the values offered by \citet{Zhang_2020}, which represent the best fit to simulation data by separate FRBs into three types: 1) One-off FRBs; 2) Repeating FRBs from spiral galaxies; 3) Repeating FRBs from dwarf galaxies.
\end{itemize}
 
The sample of 117 localized FRBs is listed in Appendix \ref{appendix_a}. For the results in Figure \ref{fig_1}b and Table \ref{tab_1}, we removed 22 FRBs for the following reasons:
\begin{itemize}
\item FRB~20221027A: More than one candidate as host galaxy.
\item FRB~20240209A, FRB~20190110C, FRB~20191106C: FRBs from elliptical galaxies, in which case $\mu_{host}$ and $\sigma_{host}$ are not provided in \citet{Zhang_2020}.
\item Other 18 FRBs: Since the prediction of $\mathrm{DM_{MW,ISM}}$ tends to fail at low Galactic latitudes, we cut $|b|$ as suggested by \citet{2020Natur.581..391M}. We use the criterion of $|b|>15^\circ$ considering the balance of sample size and systematic errors, so that 18 FRBs are removed. 
\end{itemize}

Our result $H_0=71.28^{+1.90}_{-2.08}$ km s$^{-1}$ Mpc$^{-1}$ achieves a statistical uncertainty of $\sim$2.8\%, which is the lowest among the relevant studies. Compared with recent studies, we have made the following improvements:
(1) We applied a Galactic latitude cut to reduce uncertainties in $\mathrm{DM}_{\mathrm{MW,ISM}}$ at low latitudes, which is suggested in \citet{2020Natur.581..391M};
(2) We adopted the observational constraint of $\sigma_{\mathrm{IGM}}$ from \citet{Baptista_2024}, rather than simulation-based estimates from IllustrisTNG, as recent studies (\citealt{zhang2025probingcosmicbaryondistribution}, \citealt{Medlock_2025}) have indicated that the IllustrisTNG simulation might underestimate the strength of baryonic feedback.
(3) We adopted the direction-dependent model YT2020, which is based on X-ray emission data, rather than assuming a constant value for $\mathrm{DM}_{\mathrm{MW,halo}}$, as recent studies have indicated that this component exhibits anisotropy in both observational data and cosmological simulations (\citealt{Wang_2025_universe}, \citealt{Huang_2025}).

These choices may influence the inferred value of $H_0$, and potentially explain the discrepancies seen in some previous studies. We provide a detailed investigation of these systematic effects in the following section.

\begin{deluxetable*}{ccccccc}
\tablenum{1}
\tablecaption{Comparison of parameter settings and results with recent studies.\label{tab_1}}
\tablewidth{0pt}
\tablehead{
\colhead{} & 
\colhead{\citet{Fortunato_2023}} &
\colhead{\citet{kalita2024fastradioburstsprobes}} &
\colhead{\citet{gao2024measuringhubbleconstantusing}} & \colhead{\citet{wang2025probingcosmology92localized}} & \colhead{This work}
}
\startdata
\multicolumn{6}{c}{Settings} \\
$N_{\mathrm{FRB}}$ & 23 & 64 & 69 & 88 for NE2001 & 95 for NE2001 \\
 & & & & 86 for YMW16 & 95 for YMW16\\
$f_{\mathrm{IGM}}$ & 0.82 & 0.84 & 0.93 & 0.84 & 0.93 \\
$\sigma_{\mathrm{IGM}}$ & $F=0.32$ & table from & table from & table from & $F=10^{-0.75}$ \\
 & & \citet{Zhang_2021} & \citet{Zhang_2021} & \citet{Zhang_2021} & \\
$\mu_{\mathrm{host}}$,$\sigma_{\mathrm{host}}$ & $\mu_{\mathrm{host}}=\ln(100)$ & table from & table from & table from & table from \\
 & $\sigma_{\mathrm{host}}=0.434$ & \citet{Zhang_2020} & \citet{Zhang_2020} & \citet{Zhang_2020} & \citet{Zhang_2020} \\
$\mathrm{DM_{MW,ISM}}$ & NE2001 & NE2001 & NE2001 & NE2001, YMW16 & NE2001, YMW16 \\
$\mathrm{DM_{MW,halo}}$ & 50 & $\mathcal{U}(50, 80)$ & 65 & model from & YT2020 \\
 & & & \multicolumn{3}{c}{\citet{2019MNRAS.485..648P}} \\
Galactic & no cut & no cut & no cut & no cut & $|b|>15^\circ$ \\
latitude & & & & & \\
\hline
\multicolumn{6}{c}{Result}\\
$H_0$ & $69.4\pm4.7$ & $83.53^{+2.32}_{-2.92}$ & $70.41^{+2.28}_{-2.34}$ & NE2001: $69.04^{+2.30}_{-2.07}$ & NE2001: $71.28^{+1.90}_{-2.08}$\\
& & & & YMW16: $75.61^{+2.23}_{-2.07}$ & YMW16: $71.95^{+2.03}_{-1.93}$\\
\enddata
\tablecomments{The units for $\mathrm{DM}$ and $H_0$ are pc cm$^{-3}$ and km s$^{-1}$ Mpc$^{-1}$, respectively.}
\end{deluxetable*}

\section{Discussions}
\subsection{Outliers} \label{Outliers}
We first check the extragalactic contribution of each FRB, which can be defined by the following.

\begin{equation}
\mathrm{DM_{EG}} = \mathrm{DM} - \mathrm{DM_{MW,halo}} - \mathrm{DM_{MW,ISM}}.
\label{equ_9}
\end{equation}

\begin{figure*}
\gridline{\fig{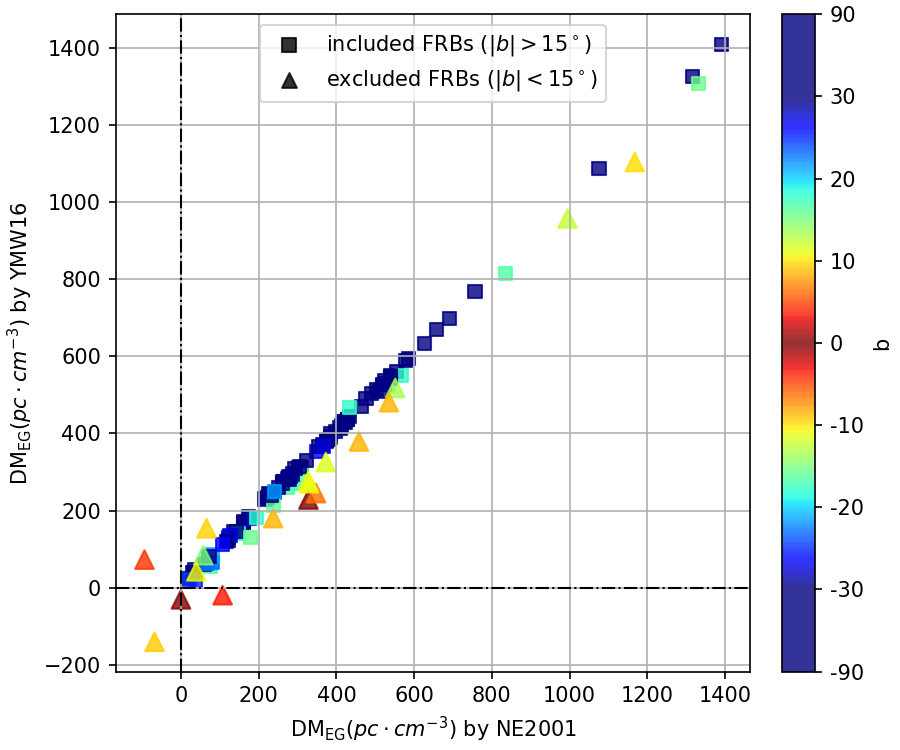}{0.33\textwidth}{(a)}
          \fig{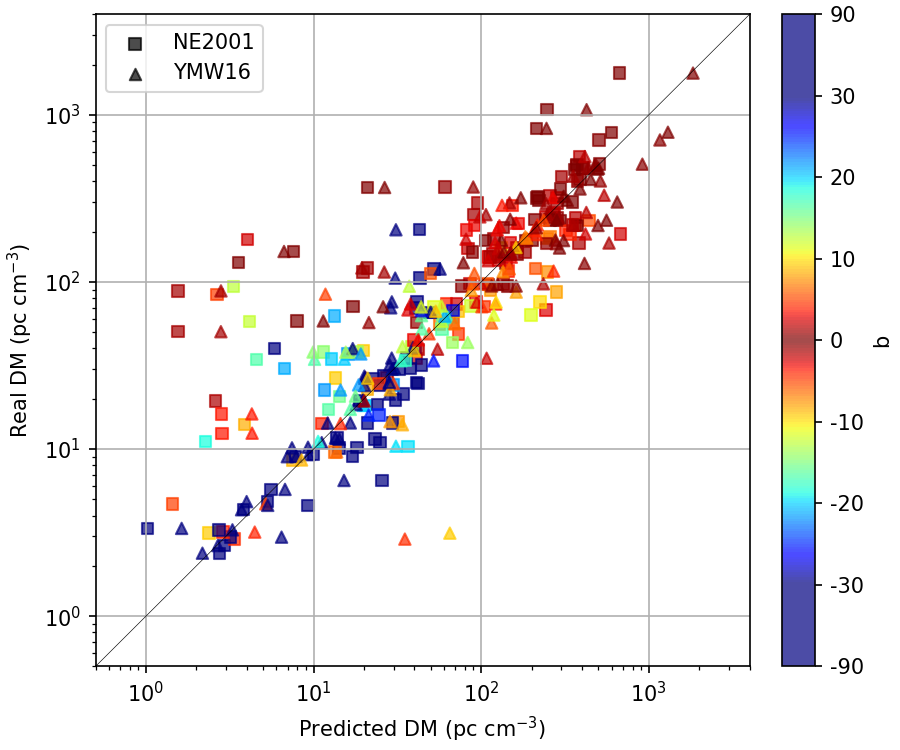}{0.33\textwidth}{(b)}
          \fig{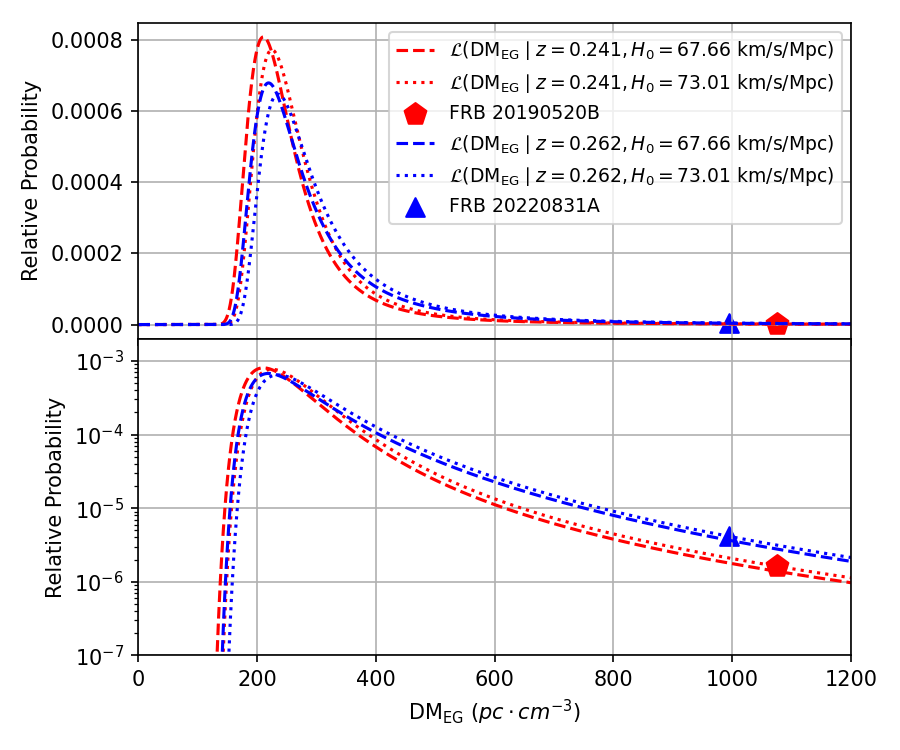}{0.33\textwidth}{(c)}
          }
\caption{(a) The $\mathrm{DM_{EG}}$ of the 117 FRBs. The $\mathrm{DM_{MW,halo}}$ is calculated with YT2020 model, and $\mathrm{DM_{MW,ISM}}$ is from NE2001 and YMW16, the colorbar refers to Galactic latitudes. The cut of $|b|>15^\circ$ can naturally remove the FRBs with negative $\mathrm{DM_{EG}}$, as well as those with large differences between NE2001 and YMW16. (b) The predicted DMs and real DMs of 189 pulsars with independently determined distances from ATNF v1.54. The colorbar refers to Galactic latitudes. The prediction tends to fail at low Galactic latitudes for both models. (c) $\mathcal{L}(\mathrm{DM_{EG}})$ calculated from Equation \ref{equ_8} for FRB~20190520B and FRB~20220831A, as well as their $\mathrm{DM_{EG}}$ values. FRBs with large DM lie in the tail and therefore have less impact on the inferred value of $H_0$.}
\label{fig_2}
\end{figure*}

The $\mathrm{DM_{EG}}$ of the 117 FRBs are plotted in Figure \ref{fig_2}a, while the $\mathrm{DM_{MW,halo}}$ is calculated with the YT2020 model, and $\mathrm{DM_{MW,ISM}}$ is from NE2001 and YMW16. We find that FRB~20210405I, FRB~20220319D, FRB~20230718A have negative $\mathrm{DM_{EG}}$ by subtracting $\mathrm{DM_{MW,ISM}}$ by NE2001, while FRB~20180916B, FRB~20220319D, FRB~20230718A are the negative ones in case of YMW16. 
Negative $\mathrm{DM_{EG}}$ results from uncertainties in the Galactic electron density model and the Galactic halo model at low Galactic latitudes, which cause $\mathcal{L}_i=0$, thus those FRBs are removed as outliers in previous studies (\citealt{kalita2024fastradioburstsprobes}, \citealt{gao2024measuringhubbleconstantusing}, \citealt{wang2025probingcosmology92localized}). To avoid selective data removal, we did not individually exclude these FRBs, but found that a uniform cut of $|b|>15^\circ$ can naturally remove those FRBs with negative $\mathrm{DM_{EG}}$. Note that cutting $|b|$ is also suggested by \citet{2020Natur.581..391M}.

After removing 18 FRBs by $|b|>15^\circ$, we discuss the influence of three special FRBs in the remaining sample. 

FRB~20190520B and FRB~20220831A deviate significantly from the Macquart relation, as can be seen from Figure \ref{fig_1}a. In particular, FRB~20190520B is a well-studied exceptional case. Its host galaxy shows a high star formation rate (SFR), and the presence of a compact persistent radio source (PRS) and significant pulse scattering together suggest that the burst originates from a highly ionized and turbulent local environment \citep{Niu_2022}. These two FRBs were excluded as outliers by \citet{gao2024measuringhubbleconstantusing}; however, the original motivation for adopting a log-normal distribution for $\mathrm{DM}_{\mathrm{host}}$ in \citet{2020Natur.581..391M} was precisely to account for cases with exceptionally large host contributions. 

Besides, FRB~20221027A was excluded by both \citet{gao2024measuringhubbleconstantusing} and \citet{wang2025probingcosmology92localized} due to ambiguity in its host galaxy localization. 

To ensure rigorous data handling and avoid selective data removal, we excluded FRB~20221027A from our final sample due to its poor localization accuracy, but retained FRB~20190520B despite its exceptionally large DM values. We note that FRB~20220831A was excluded not because of its high DM, but due to the uniform cut of $|b|>15^\circ$.

To further assess whether these special FRBs have a significant impact on $H_0$, we conducted the following comparative experiments based on our final sample of 95 FRBs:
\begin{itemize}
\item Adding FRB~20221027A into the 95 FRBs sample (i.e., 96 FRBs) or not.
\item Adding FRB~20220831A into the 95 FRBs sample (i.e., 96 FRBs) or not.
\item Removing FRB~20190520B from the 95 FRBs sample (i.e., 94 FRBs) or not.
\end{itemize}

The corresponding results are listed in Table \ref{tab_2}. 
Our findings indicate that the FRB~20190520B or FRB~20220831A results in a difference less than $0.1\sigma$, while FRB~20221027A causes a change of $\sim0.8\sigma$ and $\sim0.7\sigma$ for cases of NE2001 and YMW16, respectively.

This outcome arises because the likelihood in Equation \ref{equ_8} is composed of $p_{\mathrm{IGM}}$ and $p_{\mathrm{host}}$, each of which exhibits long-tailed behavior.
As a result, FRBs with exceptionally large DM values fall into the tail of the distribution and have less influence on the inferred value of $H_0$ (see Figure \ref{fig_2}c).
On the other hand, as shown in Figure \ref{fig_1}a, FRB~20221027A lies closer to the dense region of the sample, and therefore a poorly localized FRB like this can introduce a noticeable bias in the inferred value of $H_0$.

Therefore, under the current likelihood framework, even adding a few more high-DM FRBs to the sample is expected to have a limited impact on the inferred $H_0$; however, more careful quality control is needed for FRBs located in the dense region of the distribution.

\begin{deluxetable*}{lcc}
\tablenum{2}
\tablecaption{Comparison of $H_0$ when including the outliers or not.\label{tab_2}}
\tablewidth{0pt}
\tablehead{
 & \multicolumn{2}{c}{$H_0$ (km s$^{-1}$ Mpc$^{-1}$)}\\
 & by NE2001 & by YMW16
}
\startdata
96 FRBs with FRB~20221027A & $69.04^{+1.98}_{-1.86}$ & $70.15^{+1.71}_{-1.86}$ \\
96 FRBs with FRB~20220831A & $71.41^{+1.84}_{-2.14}$ & $72.25^{+1.89}_{-2.21}$  \\
94 FRBs without FRB~20190520B & $71.33^{+1.71}_{-2.27}$ & $71.91^{+1.92}_{-2.03}$ \\
\hline
baseline of 95 FRBs & $71.28^{+1.90}_{-2.08}$ & $71.95^{+2.03}_{-1.93}$\\
\enddata
\end{deluxetable*}

\subsection{Galactic electron density model}
The $\mathrm{DM_{MW,ISM}}$ term can be predicted by the Galactic electron density model NE2001 or YMW16. Previous studies have generally preferred to use NE2001, while a recent paper \citep{wang2025probingcosmology92localized} reported the results using both NE2001 and YMW16. We also report our results for both models in Table \ref{tab_1}, but unlike the findings of \citet{wang2025probingcosmology92localized}, the choice of Galactic electron density model only resulted in negligible variations in our $H_0$ results. To investigate whether the choice of the Galactic electron density model indeed affects the results or whether it is due to chance, we conducted a further detailed investigation.

The NE2001 model was built with 112 independent pulsar distances and scattering measures of 269 pulsars, while the YMW16 model used 189 pulsars with independently determined distances and DMs. We follow the same criterion as the YMW16 model \citep{Yao_2017} and select 189 pulsars from the ATNF Pulsar Catalogue (v1.54)\footnote{https://www.atnf.csiro.au/research/pulsar/psrcat/}. We plotted the predicted DMs and the real DMs of the 189 pulsars in Figure \ref{fig_2}b. Both models perform better at high Galactic latitudes, but tend to fail at low Galactic latitudes. Thus, \citet{2020Natur.581..391M} suggest a cut of $|b|>20^\circ$, and further remove those FRBs that have large differences in $\mathrm{DM_{MW,ISM}}$ predicted by NE2001 and YMW16.

\begin{figure*}
\gridline{\fig{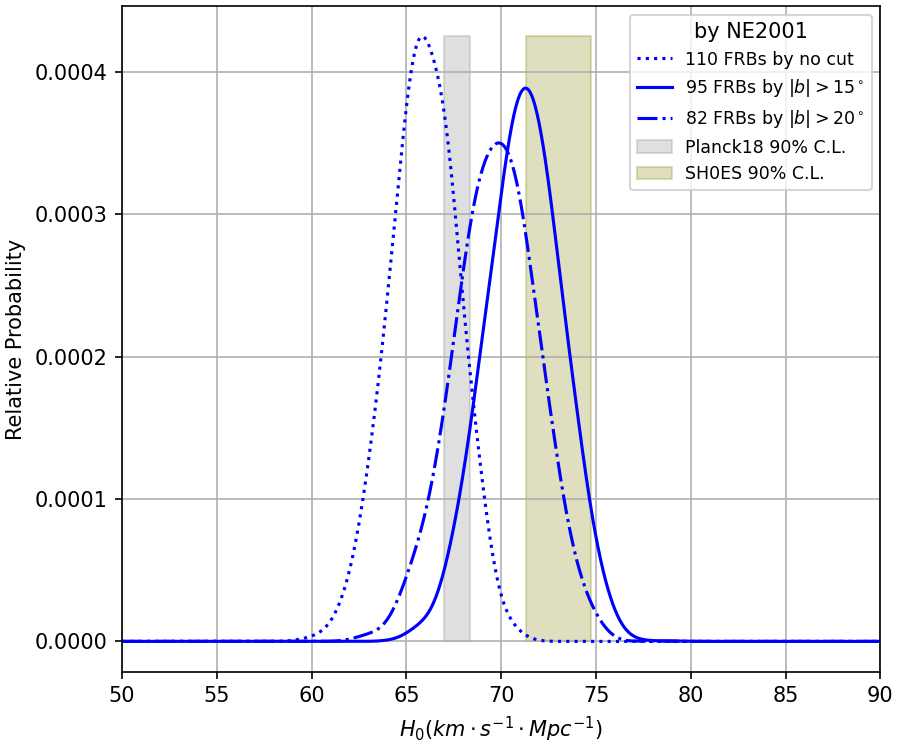}{0.48\textwidth}{(a)}
          \fig{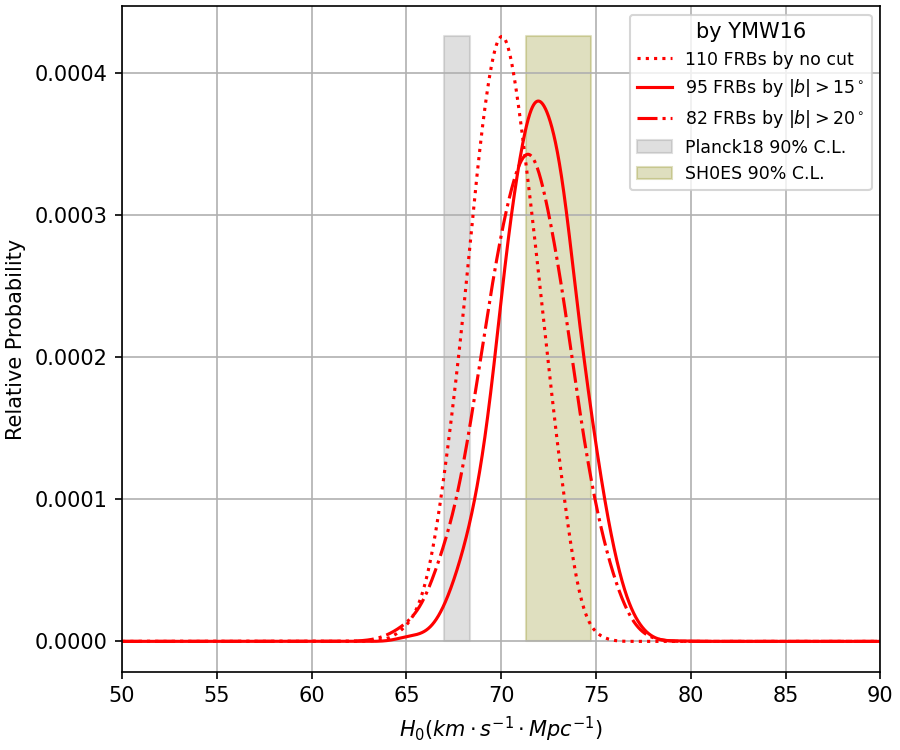}{0.48\textwidth}{(b)}
          }
\caption{(a) The $H_0$ posteriori distribution of selecting FRBs by $|b|>20^\circ$, $|b|>15^\circ$, and no $|b|$ cut. NE2001 is used as Galactic electron density model. (b) The $H_0$ posteriori distribution of selecting FRBs by $|b|>20^\circ$, $|b|>15^\circ$, and no $|b|$ cut. YMW16 is used as Galactic electron density model.}
\label{fig_3}
\end{figure*}

In Figure \ref{fig_2}a, it can be confirmed that a cut of $|b|>15^\circ$ is sufficient to remove FRBs with negative $\mathrm{DM_{EG}}$, as well as those with large differences between NE2001 and YMW16. For further investigation, we tested the impact of different $|b|$ cut criteria on the inferred value of $H_0$. The results are summarized in Table \ref{tab_3}, and a subset is visualized in Figure \ref{fig_3}. Note that when there is no criterion on $|b|$, we still have to remove the three FRBs with negative $\mathrm{DM_{EG}}$ as discussed in Section \ref{Outliers}.

As shown in Table~\ref{tab_3}, the difference between the maximum and minimum $H_0$ values under the NE2001 model is $\sim 2.1\sigma$, while for the YMW16 model the variation is $\sim 1.3\sigma$. In both cases, the results exhibit convergence after applying a certain $|b|$ cut, indicating the necessity of cutting $|b|$ as suggested by \citet{2020Natur.581..391M}.
Besides, when there is no criterion on $|b|$, the difference of $H_0$ between the case of NE2001 and YMW16 is $\sim1.6\sigma$, which is consistent with \citet{wang2025probingcosmology92localized}.

Although YMW16 appears to offer better stability, we ultimately adopted the NE2001 model with a cut of $|b|>15^\circ$ as our final results, based on the following considerations:
(1) NE2001 is used as the default model in relevant studies, including $\sigma_{\mathrm{IGM}}$ measurements \citep{Baptista_2024} and FRB localization analyses (\citealt{Sharma_2024}, \citealt{amiri_2025}), and we chose it to ensure self-consistency with these works;
(2) the NE2001-based results exhibit convergence beyond $|b|>15^\circ$, making this cut a reasonable compromise between minimizing systematic uncertainties and retaining sufficient sample size.
In the remaining sections of this paper, we use NE2001 with $|b|>15^\circ$ as the default setting.

\begin{deluxetable*}{lcc}
\tablenum{3}
\tablecaption{$H_0$ when selecting FRBs with different cut criteria on Galactic latitude.\label{tab_3}}
\tablewidth{0pt}
\tablehead{
\colhead{} & \multicolumn{2}{c}{$H_0$ (km s$^{-1}$ Mpc$^{-1}$)} \\
 & by NE2001 & by YMW16
}
\startdata
110 FRBs (no $|b|$ cut) & $65.84^{+1.85}_{-1.70}$ & $70.03^{+1.85}_{-1.78}$ \\
102 FRBs ($|b|>10^\circ$) & $65.60^{+1.83}_{-1.80}$ & $71.87^{+1.77}_{-1.98}$ \\
95 FRBs ($|b|>15^\circ$) & $71.28^{+1.90}_{-2.08}$ & $71.95^{+2.03}_{-1.93}$ \\
82 FRBs ($|b|>20^\circ$) & $69.87^{+2.01}_{-2.29}$ & $71.41^{+2.06}_{-2.42}$\\
77 FRBs ($|b|>25^\circ$) & $70.79^{+2.04}_{-2.28}$ & $73.45^{+2.09}_{-2.25}$\\
67 FRBs ($|b|>30^\circ$) & $69.95^{+2.30}_{-2.37}$ & $73.99^{+2.25}_{-2.46}$\\
\enddata
\end{deluxetable*}

\subsection{Galactic halo}
The precise value of the Galactic halo contribution, $\mathrm{DM}_{\mathrm{MW,halo}}$, remains uncertain. \citet{2019MNRAS.485..648P} modeled the halo as a spherical structure and suggested a range of $50$–$80\,\mathrm{pc\,cm^{-3}}$. Based on this assumption, several subsequent studies adopted fixed values such as $50\,\mathrm{pc\,cm^{-3}}$ (\citealt{2020Natur.581..391M}, \citealt{10.1093/mnras/stac2524}, \citealt{Fortunato_2023}) or $65\,\mathrm{pc\,cm^{-3}}$ (\citealt{Liu_2023}, \citealt{gao2024measuringhubbleconstantusing}).

However, the study of FRB~20220319D has placed upper limits on $\mathrm{DM}_{\mathrm{MW,halo}}$ along specific sightlines, with values of $28.7$ and $47.3\,\mathrm{pc\,cm^{-3}}$ derived from two nearby pulsars \citep{Ravi_2023}. These results potentially rule out the model by \citet{2019MNRAS.485..648P}, as indicated in \citet{Huang_2025}.

\begin{figure*}
\gridline{\fig{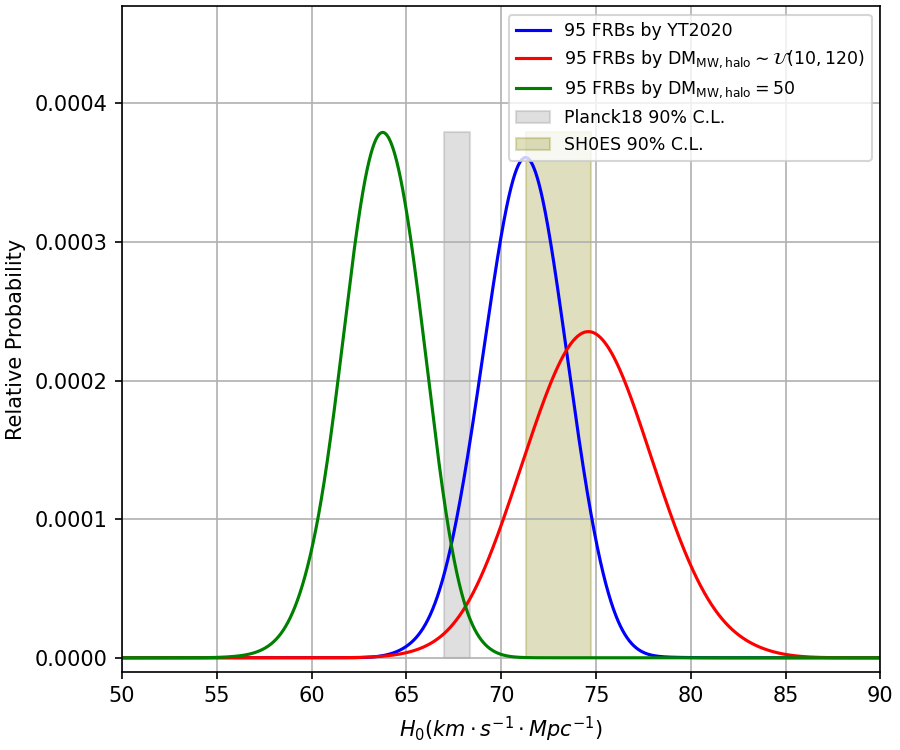}{0.33\textwidth}{(a)}
          \fig{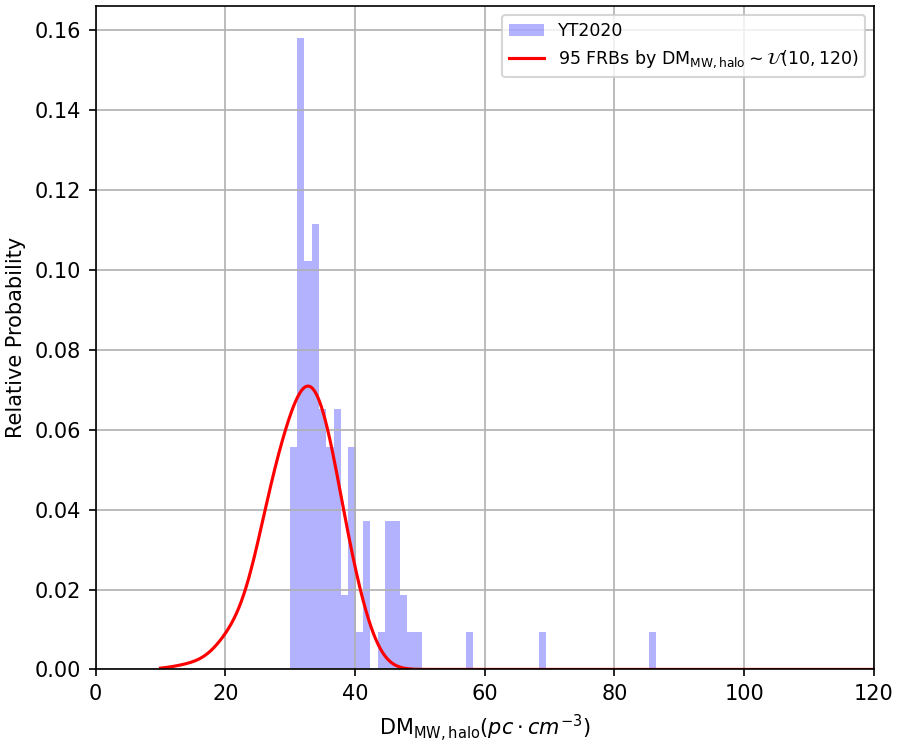}{0.33\textwidth}{(b)}
          \fig{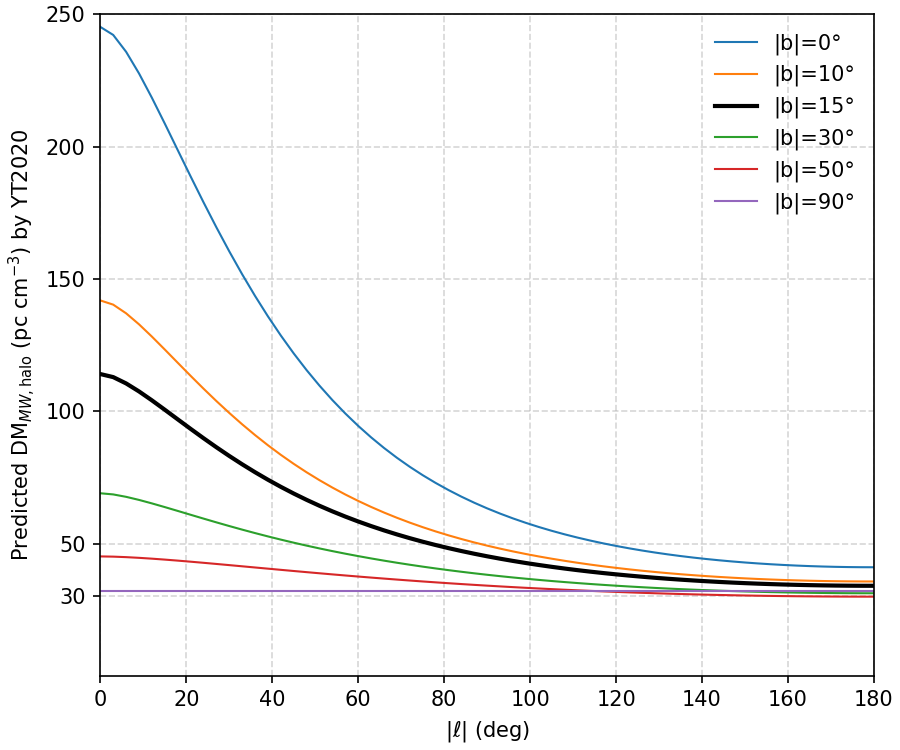}{0.33\textwidth}{(c)}
          }
\caption{(a) The $H_0$ posteriori distribution by using different values for $\mathrm{DM_{MW,halo}}$. (b) The $\mathrm{DM_{MW,halo}}$ posteriori distribution by using $\mathcal{U}(10, 120)$ as prior, and the histogram of predicted $\mathrm{DM_{MW,halo}}$ by YT2020. (c) The direction-dependence of predicted $\mathrm{DM_{MW,halo}}$ by YT2020.}
\label{fig_4}
\end{figure*}

On the other hand, \citet{Yamasaki_2020} analyzed the X-ray emission data and found that a spherical model fails to match observations. They proposed a new model YT2020 consisting of both a spherical component and a disk-like structure, which introduces a much stronger directional dependence on the predicted $\mathrm{DM}_{\mathrm{MW,halo}}$, with values ranging from $30$ to $245\,\mathrm{pc\,cm^{-3}}$ across the sky (see Figure \ref{fig_4}c). In recent FRB localization studies (\citealt{Bhardwaj_2024}, \citealt{amiri_2025}), the same fiducial value $30\,\mathrm{pc\,cm^{-3}}$ or the prediction of YT2020 is adopted.

The same trend of $\mathrm{DM}_{\mathrm{MW,halo}}$ varying with Galactic latitude has been demonstrated in both recent simulation-based and observational studies (\citealt{Huang_2025}, \citealt{Wang_2025_universe}). The YT2020 model also predicts a dependence of $\mathrm{DM}_{\mathrm{MW,halo}}$ on the Galactic longitude, with elevated values toward the Galactic center ($\ell = 0^\circ$), attributed to asymmetries in both the path length and the electron density through the halo. A comparable variation with Galactic longitude is also seen in the simulations of \citet{Huang_2025}, although no definitive explanation is provided.
Taken together, we suggest that assuming a constant value for $\mathrm{DM}_{\mathrm{MW,halo}}$ is becoming less justified. We adopt the YT2020 model for our analysis, as it emphasizes the direction-dependent structure more explicitly than \citet{2019MNRAS.485..648P} and is grounded in observational X-ray data. Nevertheless, a constant model has not been conclusively ruled out, so we also test a scenario in which $\mathrm{DM}_{\mathrm{MW,halo}}$ is treated as a fixed but unknown value.

In \citet{Cook_2023}, a constant model for $\mathrm{DM}_{\mathrm{MW,halo}}$ was constrained using a FRB sample of $|b|>30^\circ$, yielding an upper limit of $52\,\mathrm{pc\,cm^{-3}}$. When allowing for latitude dependence via a LOWESS fit, the inferred upper bounds range in \citet{Cook_2023} is from $52$ to $111\,\mathrm{pc\,cm^{-3}}$. Meanwhile, \citet{Huang_2025} reported a median value of $\mathrm{DM}_{\mathrm{MW,halo}} = 41\,\mathrm{pc\,cm^{-3}}$, with a standard deviation of $17\,\mathrm{pc\,cm^{-3}}$ and a range spanning from $13$ to $166\,\mathrm{pc\,cm^{-3}}$ across the sky, while under our $|b| > 15^\circ$ cut, the corresponding upper bound is approximately $120\,\mathrm{pc\,cm^{-3}}$. For the YT2020 model, Figure \ref{fig_4}c shows that the maximum value under our $|b|>15^\circ$ cut is likewise around $120\,\mathrm{pc\,cm^{-3}}$. Based on the above considerations, we tested a constant halo model, treating $\mathrm{DM}_{\mathrm{MW,halo}}$ as a free parameter with a uniform prior $\mathcal{U}(10, 120)\,\mathrm{pc\,cm^{-3}}$, and fit it along with $H_0$ to our final sample of 95 FRBs.

The results for $H_0$ are summarized in Table \ref{tab_4} and Figure \ref{fig_4}a. Figure \ref{fig_4}b shows the posterior distribution of $\mathrm{DM}_{\mathrm{MW,halo}}$ under the constant-halo model, which remains well within the prior bounds and is compared with the predicted values by YT2020. We also provide the result assuming $\mathrm{DM}_{\mathrm{MW,halo}} = 50\,\mathrm{pc\,cm^{-3}}$ to follow the original choice in \citet{2020Natur.581..391M}. Compared to our final result, fixing $\mathrm{DM}_{\mathrm{MW,halo}} = 50\,\mathrm{pc\,cm^{-3}}$ leads to a decrease in $H_0$ by more than $2.7\sigma$, while treating $\mathrm{DM}_{\mathrm{MW,halo}}$ as a free parameter with a uniform prior $\mathcal{U}(10, 120)$ increases $H_0$ by $\sim0.9\sigma$.

The posterior distribution of $\mathrm{DM}_{\mathrm{MW,halo}}$ shown in Figure \ref{fig_4}b peaks at $32.7\,\mathrm{pc\,cm^{-3}}$, sharing a close peak value as the YT2020 model prediction for our FRB sample. From this posterior, we also derive a 99.7\% credible upper limit of $\mathrm{DM}_{\mathrm{MW,halo}} < 43.4\,\mathrm{pc\,cm^{-3}}$ for the constant-halo model under the $|b|>15^\circ$ selection.

While we are not yet in a position to conclude that the YT2020 model is preferable to a constant-halo model, we suggest that the adoption of a constant $\mathrm{DM}_{\mathrm{MW,halo}}$ should be informed by existing observational upper limits and accompanied by a reasonable $|b|$ cut. A too large value for $\mathrm{DM}_{\mathrm{MW,halo}}$ reduces the allowable contribution from $\mathrm{DM}_{\mathrm{IGM}}$ and will consequently leads to a lower inferred value of $H_0$.

Looking ahead, we expect that a growing number of well-localized FRBs will enable the construction of more refined models of the Galactic halo.

\begin{deluxetable*}{lc}
\tablenum{4}
\tablecaption{$H_0$ and $\mathrm{DM_{MW,halo}}$ when using different values for $\mathrm{DM_{MW,halo}}$.\label{tab_4}}
\tablewidth{0pt}
\tablehead{
\colhead{} & 
\colhead{$H_0$ (km s$^{-1}$ Mpc$^{-1}$)} 
}
\startdata
95 FRBs by YT2020 & $71.28^{+1.90}_{-2.08}$ \\
95 FRBs by $\mathrm{DM_{MW,halo}}=50$ pc cm$^{-3}$ & $63.75^{+1.83}_{-1.90}$ \\
95 FRBs by using $\mathcal{U}(10, 120)$ for $\mathrm{DM_{MW,halo}}$ & $74.59^{+3.06}_{-3.06}$ \\
\enddata
\end{deluxetable*}

\subsection{Other parameter settings}
We discuss more about $f_{\mathrm{IGM}}$, $F$, $\mu_{\mathrm{host}}$, and $\sigma_{\mathrm{host}}$ which are essential in constructing the likelihood function.

For $f_{\mathrm{IGM}}$, Equation \ref{equ_3} shows that it scales $\mathrm{DM_{IGM}}$ without affecting $\mathrm{DM_{host}}$ or the MCMC sampling process. That is, if we adopt the value of $f_{\mathrm{IGM}} = 0.865$ reported by \citet{zhang2025probingcosmicbaryondistribution} instead of $f_{\mathrm{IGM}} = 0.93$ from \citet{connor2024gasrichcosmicweb}, our results would be scaled by a factor of $(0.93/0.865)$.

For $\mu_{\mathrm{host}}$ and $\sigma_{\mathrm{host}}$, \citet{Zhang_2020} fitted the IllustrisTNG simulation data and provided parameter estimates for different types of FRBs, while \citet{Fortunato_2023} considered both normal and log-normal distributions for $\mathrm{DM}_{\mathrm{host}}$, each with a median (or mean, in the normal case) of 100 and a standard deviation of 50. In our comparison, we adopt only their log-normal case. Although \citet{Fortunato_2023} did not explicitly specify the log-normal parameters, this choice corresponds to $\mu_{\mathrm{host}} = \ln(100)$ and $\sigma_{\mathrm{host}} = 0.434$. To assess the impact of these assumptions, we compare the two settings using the 85 one-off FRBs selected from our final sample of 95 events, in order to focus on a single population. In this comparison, we change only the values of $\mu_{\mathrm{host}}$ and $\sigma_{\mathrm{host}}$, which determine the assumed distribution of $\mathrm{DM}_{\mathrm{host}}$, while keeping all other modeling choices fixed and fitting $H_0$ in both cases. 

\begin{figure*}
\gridline{\fig{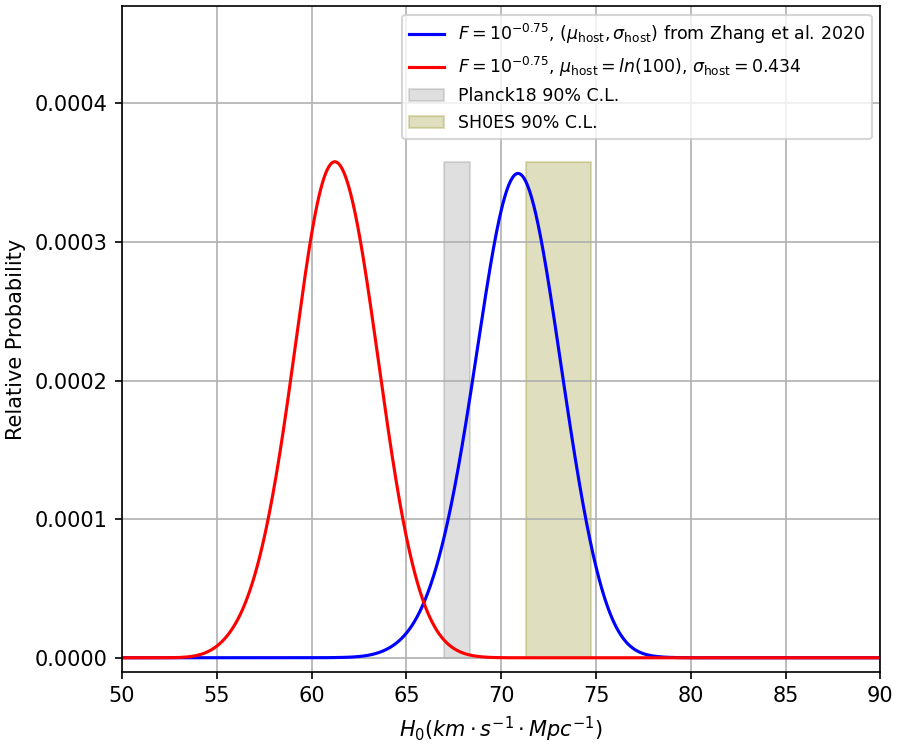}{0.48\textwidth}{(a)}
          \fig{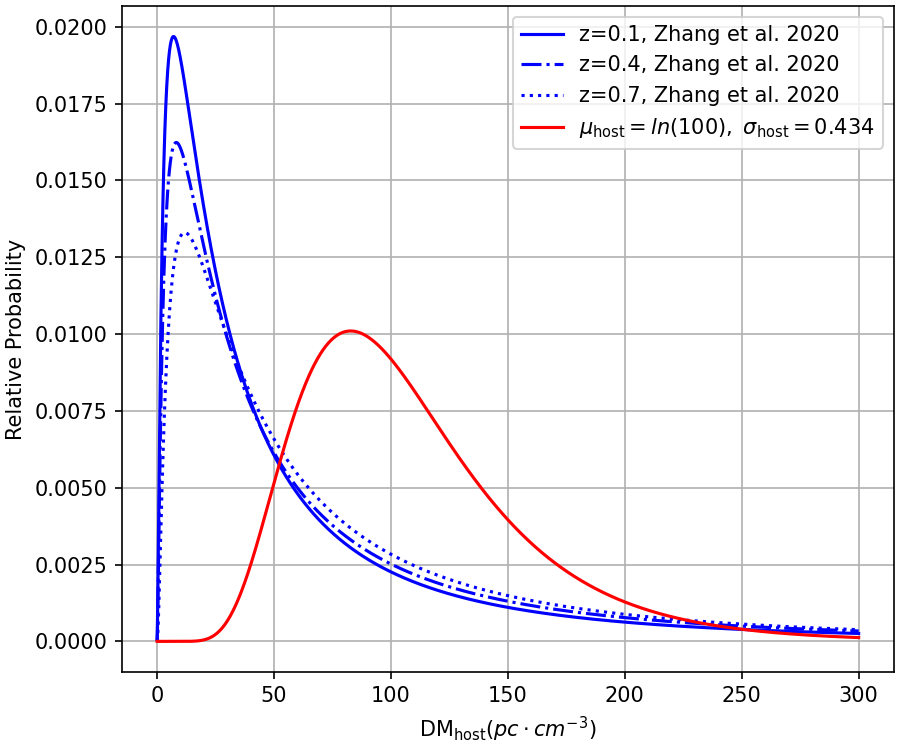}{0.48\textwidth}{(b)}
          }
\caption{(a) The $H_0$ posteriori distribution when using different $\mu_{\mathrm{host}}$ and $\sigma_{\mathrm{host}}$ on 85 one-off FRBs. The blue line use the $\mu_{\mathrm{host}}$ and $\sigma_{\mathrm{host}}$ provided by \citet{Zhang_2020}, which are sightly dependent on redshift; the red line use $\mu_{\mathrm{host}}=\ln(100)$ and $\sigma_{\mathrm{host}}=0.434$, which are assigned in \citet{Fortunato_2023}. (b) The PDF of $\mathrm{DM_{host}}$ when using different $\mu_{\mathrm{host}}$ and $\sigma_{\mathrm{host}}$. The blue lines use the values from \citet{Zhang_2020}, by $z=0.1$, $z=0.4$, $z=0.7$, respectively; the red line use the value of \citet{Fortunato_2023}.}
\label{fig_5}
\end{figure*}

The corresponding probability density functions (PDFs) for $\mathrm{DM}_{\mathrm{host}}$ are plotted in Figure~\ref{fig_5}b, and the resulting $H_0$ values are summarized in Table~\ref{tab_5}. Note that the $\mu_{\mathrm{host}}$ and $\sigma_{\mathrm{host}}$ values from \citet{Zhang_2020} exhibit a mild dependence on redshift, primarily due to the increasing local electron density in high-redshift galaxies and the $(1+z)$ scaling applied when converting intrinsic $\mathrm{DM}_{\mathrm{host}}$ to observed values. The \citet{Fortunato_2023} setting results in a systematically larger $\mathrm{DM}_{\mathrm{host}}$ distribution, which in turn reduces the inferred contribution from $\mathrm{DM}_{\mathrm{IGM}}$ and consequently leads to a decrease in $H_0$ by more than 3$\sigma$.

For parameter $F$, \citet{2020Natur.581..391M} explored values in the full range of $[0, 0.5]$, but ultimately recommended a physically motivated subrange of $[0.09, 0.32]$ based on semi-analytic models for marginalization. This range corresponds to plausible baryonic feedback scenarios, with $F=0.09, 0.15, 0.32$ representing strong to weak feedback, respectively. Subsequent studies have generally adopted the value $F=0.32$ (\citealt{James_2021}, \citealt{10.1093/mnras/stac2524}, \citealt{Fortunato_2023}, \citealt{Moroianu_2023}). In more recent work, \citet{Baptista_2024} reported a best-fit value of $F=10^{-0.75}$, suggesting that earlier analyses based on $F=0.32$ may have underestimated the strength of baryonic feedback. On the other hand, several recent studies adopt the IGM variance model fitted from the IllustrisTNG simulation, while \citet{zhang2025probingcosmicbaryondistribution} and \citet{Medlock_2025} indicated that the IllustrisTNG simulation might also underestimate the strength of baryonic feedback.

In our main analysis, we adopt the latest result of $F=10^{-0.75}$ from \citet{Baptista_2024}. To assess how varying baryonic feedback strength impacts the inferred $H_0$, we also test the two cases of $F=0.09$ and $F=0.32$. For simplicity, this comparison uses the same one-off population of 85 FRBs selected from our final 95-FRB sample, and all other modeling choices are kept fixed as described in Table~\ref{tab_1}. 

\begin{figure*}
\gridline{\fig{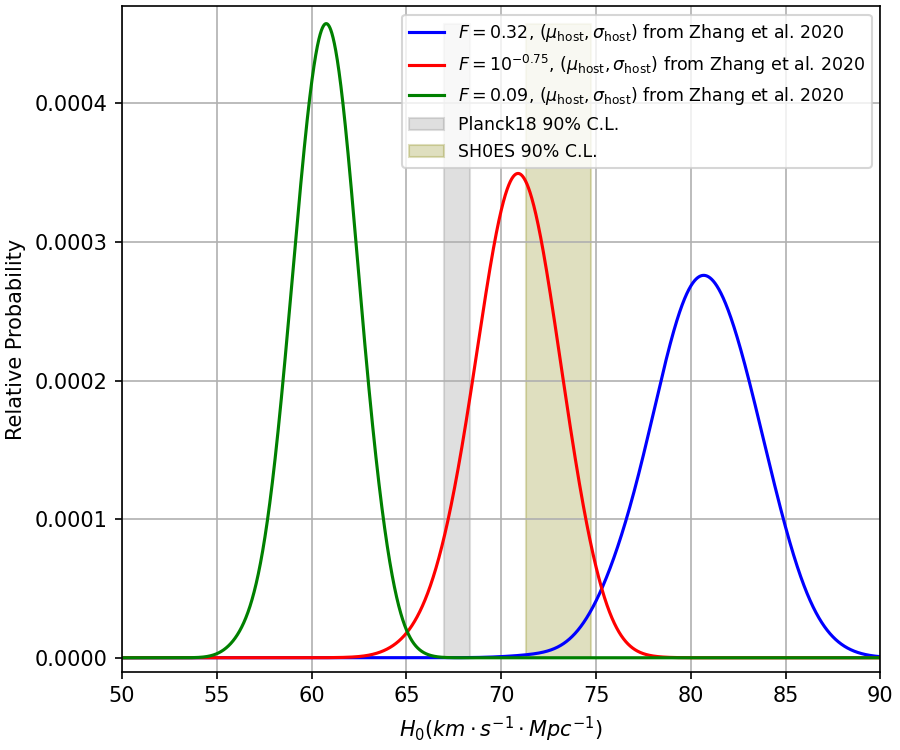}{0.48\textwidth}{(a)}
          \fig{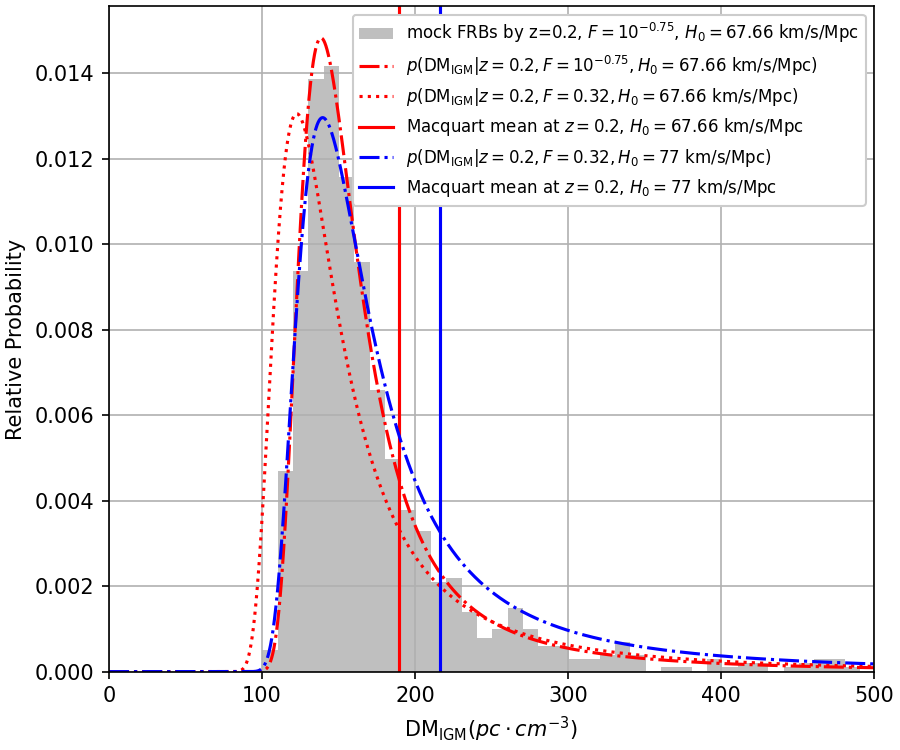}{0.48\textwidth}{(b)}
          }
\caption{(a) Posteriori of $H_0$ posteriori for three different values of $F$ on 85 one-off FRBs, showing that larger $F$ shifts $H_0$ higher. (b) Probability density functions of $\mathrm{DM_{IGM}}$ at $z=0.2$, illustrating the degeneracy between $H_0$ and $F$: $H_0$ shifts the distribution horizontally, while $F$ controls its width and peak location.}
\label{fig_6}
\end{figure*}

Our results are summarized in Table~\ref{tab_5} and Figure~\ref{fig_6}a. We find that increasing $F$ from $10^{-0.75}$ to $0.32$ raises the inferred $H_0$ by $\sim2.9\sigma$, while decreasing $F$ from $10^{-0.75}$ to $0.09$ lowers $H_0$ by $\sim3.9\sigma$. These results demonstrate that the choice of $F$ has a critical impact on the inferred value of the Hubble constant. A strong coupling between $H_0$ and baryonic feedback, as parameterized by $F$, was also identified by \citet{Baptista_2024}, but has often been overlooked in earlier studies.

To visualize the origin of this coupling, we illustrate in Figure~\ref{fig_6}b the PDF of $\mathrm{DM}_{\mathrm{IGM}}$ at a fixed redshift with Equation~\ref{equ_4}. Numerically, $H_0$ primarily controls the mean value, while $F$ determines the standard deviation. However, as the PDF is skewed, its mean and peak do not coincide. When $H_0$ is fixed, increasing $F$ flattens the distribution and slightly raises the tail; to satisfy the requirement $\langle \Delta \rangle = 1$ in Equation~\ref{equ_4}, this results in a shift of the distribution toward lower $\mathrm{DM}_{\mathrm{IGM}}$. In contrast, increasing $H_0$ alone shifts the entire distribution toward higher $\mathrm{DM}_{\mathrm{IGM}}$ without changing its shape. This leads to a degeneracy between the two parameters.

Briefly, the parameters $f_{\mathrm{IGM}}$, $F$, $\mu_{\mathrm{host}}$, and $\sigma_{\mathrm{host}}$ all require careful treatment. In particular, earlier studies may have overestimated the value of $F$ (i.e., underestimating the true level of baryonic feedback), leading to a bias toward higher $H_0$. While for $\mathrm{DM_{host}}$, the commonly used log-normal form is introduced by \citet{2020Natur.581..391M} empirically, and recent work \citep{zhang2025probingcosmicbaryondistribution} has begun to refine this treatment. We expect future data to provide stronger constraints on these key parameters.

\begin{deluxetable*}{ccc}
\tablenum{5}
\tablecaption{$H_0$ when applying different $F$, $\mu_{\mathrm{host}}$ and $\sigma_{\mathrm{host}}$ with 85 one-off FRBs.\label{tab_5}}
\tablewidth{0pt}
\tablehead{
\colhead{$F$} & 
\colhead{$\mu_{\mathrm{host}}$, $\sigma_{\mathrm{host}}$} &
\colhead{$H_0$ (km s$^{-1}$ Mpc$^{-1}$)}
}
\startdata
$F=10^{-0.75}$ & from \citet{Zhang_2020} & $70.89^{+1.96}_{-2.13}$ \\
$F=10^{-0.75}$ & $\mu_{\mathrm{host}}=\ln(100)$, $\sigma_{\mathrm{host}}=0.434$ & $61.22^{+2.00}_{-1.94}$ \\
\hline
$F=0.32$ & from \citet{Zhang_2020} & $80.68^{+2.59}_{-2.56}$ \\
$F=0.09$ & from \citet{Zhang_2020} & $60.77^{+1.50}_{-1.60}$ \\
\enddata
\end{deluxetable*}

\section{Summary}
We constrain the Hubble constant to $H_0=71.28^{+1.90}_{-2.08}$ km s$^{-1}$ Mpc$^{-1}$ under the $\Lambda$CDM model, achieving a statistical uncertainty of approximately $2.8$\%, based on a subset of 95 FRBs selected from an initial sample of 117 localized events. The posterior distribution of $H_0$ exhibits mild non-Gaussianity, which we account for by quoting asymmetric error bars. As the posteriors in both our analysis and previous works do not show a consistent skew toward any particular direction, the observed asymmetry is likely attributable to sample variance rather than to a systematic feature of the likelihood function.

Beyond statistical uncertainties, we performed a comprehensive assessment of key sources of systematic error, leading to the following conclusions.

We find it is important to follow \citet{2020Natur.581..391M} in applying a Galactic latitude cut—not only to mitigate modeling uncertainties in the Galactic electron density at low latitudes, but also to naturally exclude events with unphysical negative extragalactic dispersion measures. In our case, we adopted a cut of $|b|>15^\circ$ based on the NE2001 model and confirmed that the results remain stable under stricter latitude selections.

For the Galactic halo contribution, we do not adopt a fixed value as in some previous studies, but instead use the YT2020 model, which provides a direction-dependent prediction. While a constant model has not been conclusively ruled out, we suggest that any fixed value be guided by known upper limits, as overly large values may bias $H_0$ toward lower values. For example, an upper limit of $52\,\mathrm{pc\,cm^{-3}}$ was reported by \citet{Cook_2023} for $|b|>30^\circ$; or, if low-latitude events are included, the constraint of $47.3\,\mathrm{pc\,cm^{-3}}$ from \citet{Ravi_2023}, based on FRB~20220319D, should be considered. In addition, we also tested a constant model in which the Galactic halo contribution is treated as a fixed but unknown value, and derived a 99.7\% credible upper limit of $43.4\,\mathrm{pc\,cm^{-3}}$ under our $|b|>15^\circ$ selection.

For the other components, the IGM fluctuation parameter $F$ and the $\mathrm{DM_{host}}$ contribution also influence the inferred value of $H_0$. In particular, we confirm a strong coupling between $F$ and $H_0$: larger values of $F$, which correspond to weaker baryonic feedback and greater IGM variance, systematically lead to higher values of $H_0$. This highlights the importance of carefully selecting or constraining $F$ in FRB-based cosmological analyses.

Despite these challenges, we anticipate that future localized FRB samples will enable more refined modeling of each component and further improve the robustness of FRB-based cosmology.

\acknowledgments
This work is supported by by the Leading Innovation and Entrepreneurship Team of Zhejiang Province of China grant No. 2023R01008, National Natural Science Foundation of China grant No.\ 12203045, and by Key R\&D Program of Zhejiang grant No. 2024SSYS0012.

%






\appendix

\section{Dataset}\label{appendix_a}

\startlongtable
\begin{deluxetable*}{lclllll}
\tablenum{6}
\tablecaption{Data sample of 117 localized FRBs\label{tab_6}}
\tablewidth{0pt}
\tablehead{
\colhead{Name} & 
\colhead{Type} &
\colhead{$\ell \; (^\circ)$} &
\colhead{$b \; (^\circ)$} & 
\colhead{DM (pc cm$^{-3}$)} & 
\colhead{Redshift} &
\colhead{Reference}
}
\startdata
FRB~20121102A & 3 & 174.89 & -0.23 & 557.0 & 0.1927 & \citet{Heintz_2020} \\
FRB~20171020A & 1 & 29.3 & -51.3 & 114.0 & 0.00867 & \citet{Mahony_2018} \\
FRB~20180301A & 3 & 204.412 & -6.481 & 536.0 & 0.3304 & \citet{Bhandari_2022} \\
FRB~20180814A & 2 & 136.46 & 16.58 & 198.4 & 0.0781 & \citet{Michilli_2023} \\
FRB~20180916B & 2 & 129.71 & 3.73 & 348.8 & 0.0337 & \citet{Heintz_2020} \\
FRB~20180924B & 1 & 0.7424 & -49.4147 & 362.16 & 0.3212 & \citet{Heintz_2020} \\
FRB~20181030A & 3 & 133.4 & 40.9 & 103.5 & 0.0039 & \citet{Bhardwaj_2021} \\
FRB~20181112A & 1 & 342.5995 & -47.6988 & 589.0 & 0.4755 & \citet{Heintz_2020} \\
FRB~20181220A & 1 & 105.24 & -10.73 & 208.7 & 0.02746 & \citet{Bhardwaj_2024}\\
FRB~20181223C & 1 & 207.75 & 79.51 & 111.6 & 0.03024 & \citet{Bhardwaj_2024}\\
FRB~20190102C & 1 & 312.6537 & -33.4931 & 364.545 & 0.2913 & \citet{Heintz_2020}\\
FRB~20190110C & 4 & 65.52 & 43.85 & 221.6 & 0.12244 & \citet{Ibik_2024}\\
FRB~20190303A & 3 & 97.5 & 65.7 & 222.4 & 0.064 & \citet{Michilli_2023}\\
FRB~20190418A & 1 & 179.3 & -22.93 & 182.8 & 0.07132 & \citet{Bhardwaj_2024}\\
FRB~20190425A & 1 & 42.06 & 33.02 & 127.8 & 0.03122 & \citet{Bhardwaj_2024}\\
FRB~20190520B & 3 & 359.67 & 29.91 & 1205.0 & 0.241 & \citet{Niu_2022}\\
FRB~20190523A & 1 & 117.03 & 44.0 & 760.8 & 0.66 & \citet{Heintz_2020}\\
FRB~20190608B & 1 & 53.2088 & -48.5296 & 340.05 & 0.1178 & \citet{Heintz_2020}\\
FRB~20190611B & 1 & 312.9352 & -33.2818 & 321.0 & 0.3778 & \citet{Heintz_2020}\\
FRB~20190614D & 1 & 136.3 & 16.5 & 959.0 & 0.6 & \citet{Law_2020}\\
FRB~20190711A & 3 & 310.9081 & -33.902 & 592.6 & 0.522 & \citet{Heintz_2020}\\
FRB~20190714A & 1 & 289.6972 & 48.9359 & 504.13 & 0.2365 & \citet{Heintz_2020}\\
FRB~20191001A & 1 & 341.2267 & -44.9039 & 507.9 & 0.234 & \citet{Heintz_2020}\\
FRB~20191106C & 4 & 105.7 & 73.2 & 332.2 & 0.10775 & \citet{Ibik_2024}\\
FRB~20191228A & 1 & 20.5553 & -64.9245 & 297.5 & 0.2432 & \citet{Bhandari_2022}\\
FRB~20200120E & 4 & 142.19 & 41.22 & 87.82 & 0.0008 & \citet{Bhardwaj_2021b}\\
FRB~20200223B & 2 & 118.1 & -33.9 & 201.8 & 0.06024 & \citet{Ibik_2024}\\
FRB~20200430A & 1 & 17.1396 & 52.503 & 380.25 & 0.1608 & \citet{Heintz_2020}\\
FRB~20200906A & 1 & 202.257 & -49.9989 & 577.8 & 0.3688 & \citet{Bhandari_2022}\\
FRB~20201123A & 1 & 340.23 & -9.68 & 433.9 & 0.0507 & \citet{Rajwade_2022}\\
FRB~20201124A & 2 & 177.6 & -8.5 & 413.52 & 0.0979 & \citet{Fong_2021}\\
FRB~20210117A & 1 & 45.9175 & -57.6464 & 728.95 & 0.2145 & \citet{Gordon_2023}\\
FRB~20210320C & 1 & 318.8729 & 45.3081 & 384.8 & 0.2797 & \citet{Gordon_2023}\\
FRB~20210405I & 1 & 338.19 & -4.59 & 566.43 & 0.066 & \citet{Driessen_2023}\\
FRB~20210410D & 1 & 312.32 & -34.13 & 578.78 & 0.1415 & \citet{Gordon_2023}\\
FRB~20210603A & 1 & 119.71 & -41.58 & 500.147 & 0.1772 & \citet{Cassanelli_2024}\\
FRB~20210807D & 1 & 39.8612 & -14.8775 & 251.9 & 0.1293 & \citet{Gordon_2023}\\
FRB~20211127I & 1 & 312.0214 & 43.5427 & 243.83 & 0.0469 & \citet{Gordon_2023}\\
FRB~20211203C & 1 & 314.5185 & 30.4361 & 636.2 & 0.3437 & \citet{Gordon_2023}\\
FRB20211212A & 1 & 244.0081 & 47.3154 & 206.0 & 0.0707 & \citet{Gordon_2023}\\
FRB~20220105A & 1 & 18.555 & 74.808 & 583.0 & 0.2785 & \citet{Gordon_2023}\\
FRB~20220204A & 1 & 102.26 & 27.06 & 612.584 & 0.4012 & \citet{Sharma_2024}\\
FRB~20220207C & 1 & 106.94 & 18.39 & 262.38 & 0.0433 & \citet{Law_2024}\\
FRB~20220208A & 1 & 107.62 & 15.36 & 440.73 & 0.351 & \citet{Sharma_2024}\\
FRB~20220307B & 1 & 116.24 & 10.47 & 499.27 & 0.2481 & \citet{Law_2024}\\
FRB~20220310F & 1 & 140.02 & 34.8 & 462.24 & 0.478 & \citet{Law_2024}\\
FRB~20220319D & 1 & 129.18 & 9.11 & 110.98 & 0.0112 & \citet{Law_2024}\\
FRB~20220330D & 1 & 134.18 & 42.93 & 467.788 & 0.3714 & \citet{Sharma_2024}\\
FRB~20220418A & 1 & 110.75 & 44.47 & 623.25 & 0.6214 & \citet{Law_2024}\\
FRB~20220501C & 1 & 11.1777 & -71.4731 & 449.5 & 0.381 & \citet{shannon_2025}\\
FRB~20220506D & 1 & 108.35 & 16.51 & 396.97 & 0.3004 & \citet{Law_2024}\\
FRB~20220509G & 1 & 100.94 & 25.48 & 269.53 & 0.0894 & \citet{Law_2024}\\
FRB~20220529A & 3 & 130.79 & -41.86 & 246.0 & 0.1839 & \citet{gao2024measuringhubbleconstantusing}\\
FRB~20220610A & 1 & 8.8392 & -70.1857 & 1458.1 & 1.016 & \citet{Ryde_2023}\\
FRB~20220717A & 1 & 19.8352 & -17.632 & 637.34 & 0.3630 & \citet{Rajwade_2024}\\
FRB~20220725A & 1 & 0.0017 & -71.1863 & 290.4 & 0.1926 & \citet{shannon_2025}\\
FRB~20220726A & 1 & 139.97 & 17.57 & 686.232 & 0.3619 & \citet{Sharma_2024}\\
FRB~20220825A & 1 & 106.99 & 17.79 & 651.24 & 0.2414 & \citet{Law_2024}\\
FRB~20220831A & 1 & 110.96 & 12.47 & 1146.14 & 0.262 & \citet{Sharma_2024}\\
FRB~20220912A & 2 & 347.27 & 48.7 & 219.46 & 0.0771 & \citet{Ravi_2023}\\
FRB~20220914A & 1 & 104.31 & 26.13 & 631.29 & 0.1139 & \citet{Law_2024}\\
FRB~20220918A & 1 & 300.6851 & -46.2342 & 656.8 & 0.491 & \citet{shannon_2025}\\
FRB~20220920A & 1 & 104.92 & 38.89 & 314.99 & 0.1582 & \citet{Law_2024}\\
FRB~20221012A & 1 & 101.14 & 26.14 & 441.08 & 0.2847 & \citet{Law_2024}\\
FRB~20221022A & 1 & 124.942 & 24.576 & 116.8371 & 0.0149 & \citet{Mckinven2025}\\
FRB~20221027A & 1 & 142.66 & 33.96 & 452.723 & 0.5422 & \citet{Sharma_2024}\\
FRB~20221029A & 1 & 140.39 & 38.01 & 1391.75 & 0.975 & \citet{Sharma_2024}\\
FRB~20221101B & 1 & 113.25 & 11.06 & 491.554 & 0.2395 & \citet{Sharma_2024}\\
FRB~20221106A & 1 & 220.901 & -50.8788 & 344.0 & 0.2044 & \citet{shannon_2025}\\
FRB~20221113A & 1 & 139.53 & 16.99 & 411.027 & 0.2505 & \citet{Sharma_2024}\\
FRB~20221116A & 1 & 124.47 & 8.71 & 643.448 & 0.2764 & \citet{Sharma_2024}\\
FRB~20221219A & 1 & 103.19 & 34.07 & 706.708 & 0.553 & \citet{Sharma_2024}\\
FRB~20230124A & 1 & 107.55 & 40.25 & 590.574 & 0.0939 & \citet{Sharma_2024}\\
FRB~20230203A & 1 & 188.7125 & 54.087 & 420.1 & 0.1464 & \citet{amiri_2025}\\
FRB~20230216A & 1 & 242.6 & 46.33 & 828.289 & 0.531 & \citet{Sharma_2024}\\
FRB~20230222A & 1 & 204.7164 & 8.6957 & 706.1 & 0.1223 & \citet{amiri_2025}\\
FRB~20230222B & 1 & 49.6176 & 49.9787 & 187.8 & 0.1100 & \citet{amiri_2025}\\
FRB~20230307A & 1 & 127.35 & 45.0 & 608.854 & 0.2706 & \citet{Sharma_2024}\\
FRB~20230311A & 1 & 157.7134 & 16.0395 & 364.3 & 0.1918 & \citet{amiri_2025}\\
FRB~20230501A & 1 & 112.43 & 11.51 & 532.471 & 0.3015 & \citet{Sharma_2024}\\
FRB~20230521B & 1 & 115.65 & 9.97 & 1342.88 & 1.345 & \citet{Sharma_2024}\\
FRB~20230526A & 1 & 290.171 & -63.4721 & 316.4 & 0.157 & \citet{shannon_2025}\\
FRB~20230626A & 1 & 105.68 & 38.34 & 452.723 & 0.327 & \citet{Sharma_2024}\\
FRB~20230628A & 1 & 135.48 & 42.21 & 344.952 & 0.127 & \citet{Sharma_2024}\\
FRB~20230703A & 1 & 137.2098 & 67.4750 & 291.3 & 0.1184 & \citet{amiri_2025}\\
FRB~20230708A & 1 & 342.6288 & -33.3877 & 411.51 & 0.105 & \citet{shannon_2025}\\
FRB~20230712A & 1 & 132.31 & 43.69 & 587.567 & 0.4525 & \citet{Sharma_2024}\\
FRB~20230718A & 1 & 259.4629 & -0.3666 & 476.6 & 0.0357 & \citet{Glowacki_2024}\\
FRB~20230730A & 1 & 158.8166 & -17.8164 & 312.5 & 0.2115 & \citet{amiri_2025}\\
FRB~20230808F & 1 & 263.5216 & -50.9503 & 653.2 & 0.3472 & \citet{Hanmer_2025}\\
FRB~20230814A & 1 & 112.56 & 13.2 & 696.411 & 0.5535 & \citet{Sharma_2024}\\
FRB~20230902A & 1 & 256.9906 & -53.3387 & 440.1 & 0.3619 & \citet{shannon_2025}\\
FRB~20230926A & 1 & 68.2353 & 27.4840 & 222.8 & 0.0553 & \citet{amiri_2025}\\
FRB~20231005A & 1 & 57.1653 & 44.3532 & 189.4 & 0.0713 & \citet{amiri_2025}\\
FRB~20231011A & 1 & 127.2287 & -20.9430 & 186.3 & 0.0783 & \citet{amiri_2025}\\
FRB~20231017A & 1 & 100.6098 & -21.6519 & 344.2 & 0.2450 & \citet{amiri_2025}\\
FRB~20231025B & 1 & 93.4332 & 29.4340 & 368.7 & 0.3238 & \citet{amiri_2025}\\
FRB~20231120A & 1 & 140.43 & 37.94 & 437.737 & 0.0368 & \citet{Sharma_2024}\\
FRB~20231123A & 1 & 199.3094 & -15.7427 & 302.1 & 0.0729 & \citet{amiri_2025}\\
FRB~20231123B & 1 & 105.71 & 38.38 & 396.857 & 0.2621 & \citet{Sharma_2024}\\
FRB~20231201A & 1 & 163.0405 & -22.7702 & 169.4 & 0.1119 & \citet{amiri_2025}\\
FRB~20231206A & 1 & 161.0582 & 27.4819 & 457.7 & 0.0659 & \citet{amiri_2025}\\
FRB~20231220A & 1 & 143.29 & 31.69 & 491.2 & 0.3355 & \citet{Sharma_2024}\\
FRB~20231223C & 1 & 52.3107 & 32.0802 & 165.8 & 0.1059 & \citet{amiri_2025}\\
FRB~20231226A & 1 & 236.5768 & 48.6458 & 329.9 & 0.1569 & \citet{shannon_2025}\\
FRB~20231229A & 1 & 135.3449 & -26.4433 & 198.5 & 0.0190 & \citet{amiri_2025}\\
FRB~20231230A & 1 & 195.8701 & -25.2387 & 131.4 & 0.0298 & \citet{amiri_2025}\\
FRB~20240114A & 3 & 57.73 & -31.68 & 527.7 & 0.1306 & \citet{Tian_2024}\\
FRB~20240119A & 1 & 112.37 & 43.24 & 483.626 & 0.37 & \citet{Sharma_2024}\\
FRB~20240123A & 1 & 138.14 & 15.34 & 1462.39 & 0.968 & \citet{Sharma_2024}\\
FRB~20240201A & 1 & 222.1335 & 47.9692 & 374.5 & 0.0427 & \citet{shannon_2025}\\
FRB~20240209A & 4 & 118.5592 & 26.5812 & 176.518 & 0.1384 & \citet{Eftekhari_2025}\\
FRB~20240210A & 1 & 14.4396 & -86.2116 & 283.73 & 0.0237 & \citet{shannon_2025}\\
FRB~20240213A & 1 & 136.49 & 41.57 & 358.141 & 0.185 & \citet{Sharma_2024}\\
FRB~20240215A & 1 & 102.29 & 30.17 & 549.461 & 0.21 & \citet{Sharma_2024}\\
FRB~20240229A & 1 & 131.13 & 44.11 & 491.15 & 0.287 & \citet{Sharma_2024}\\
FRB~20240310A & 1 & 291.7066 & -72.2713 & 601.8 & 0.127 & \citet{shannon_2025}\\
\enddata
\tablecomments{The types of FRBs: 1) One-off FRBs; 2) Repeating FRBs from spiral galaxies; 3) Repeating FRBs from dwarf galaxies; 4) Repeating FRBs from the other galaxies.}
\end{deluxetable*}


\bibliography{sample63}{}
\bibliographystyle{aasjournal}



\end{document}